# Air-stable bright entangled photon-pair source from graphene-encapsulated van der Waals ferroelectric NbOI$_2$

Mayank Joshi[1,2], Mengting Jiang[3], Yu Xing[1], Yuerui Lu[4], Jie Zhao[2], Ping Koy Lam[1,3,5], Syed M Assad[1,3,5*], Xuezhi Ma[1,3,5*], and Young-Wook Cho[1,3,5*]

1. Quantum Innovation Centre (Q. InC), Agency for Science, Technology and Research (A*STAR), 4 Fusionopolis Way, Kinesis, #05, 138635, Singapore.

2. Australian National University (ANU) Department of Quantum Science and Technology, Australian National University, Canberra, ACT, 2601, Australia.

3. Institute of Materials Research and Engineering (IMRE), Agency for Science, Technology and Research, 2 Fusionopolis Way, Innovis, #08-03, Singapore, 138634, Singapore.

4. School of Engineering, College of Engineering, Computing and Cybernetics, the Australian National University, Canberra, ACT, 2601, Australia

5. Centre for Quantum Technologies (CQT), National University of Singapore (NUS), 117543, Singapore.

Corresponding authors email address:
Young-Wook Cho: cho_youngwook@a-star.edu.sg; Xuezhi Ma: ma_xuezhi@a-star.edu.sg; Syed M Assad: cqtsma@gmail.com

**Abstract**

Van der Waals (vdW) ferroelectrics are emerging nonlinear photonic materials that combine large second-order susceptibility $\chi^{(2)}$ with heterostructure compatibility, offering an attractive route toward miniaturized spontaneous parametric down-conversion (SPDC) sources. However, vdW SPDC sources operating under continuous irradiation in air remain limited in low brightness and poor operational stability, as oxygen and moisture exposure, together with pump-induced heating, lead to material degradation and permanent damage. Here we demonstrate an air-stable, bright SPDC source based on ferroelectric $NbOI_2$ enabled by graphene encapsulation. Graphene provides robust environmental protection and can effectively supress pump induced degradation by enhancing heat dissipation. We report a record photon-pair generation absolute rate of 258 Hz and a normalized brightness of 19,900 Hz/(mW·mm). Leveraging this stabilized platform, we further generate polarization entangled photon pairs with 94% fidelity with respect to the maximally entangled Bell states from graphene-encapsulated 90° twisted bilayer $NbOI_2$. Our results establish a practical and air-stable vdW ferroelectric SPDC platform that overcomes key limitations of existing vdW quantum light sources and provides a viable pathway toward scalable, integrated entangled photon sources for on chip quantum photonics.

**Main**

Spontaneous parametric down-conversion (SPDC)[1-3] is a widely used mechanism for generating correlated and entangled photon pairs. It serves as a fundamental resource across a broad range of photonic quantum technologies, including quantum communication[4], quantum computing[5], quantum sensing[6] and foundational tests of quantum physics[7]. Conventional SPDC sources based on bulk nonlinear crystals have enabled many proof of principle demonstrations, but their size and limited integrability motivate the development of compact photon pair sources compatible with emerging quantum photonic platforms.

Van der Waals (vdW) nonlinear materials provide an attractive route toward such compact SPDC sources, as they combine strong second order susceptibility ($\chi^{(2)}$) with natural compatibility for heterostructure integration[8-12]. Their dangling bond free surfaces facilitate interfacing with diverse optical, electronic, and protective layers, while thickness control and deterministic stacking enable device architectures in which nonlinear generation, environmental protection, and thermal management can be engineered within an atomically thin platform.

Despite this promise, practical vdW SPDC sources face two major challenges. First, the achievable photon pairs often remain insufficient for scalable quantum photonic applications. Second, operational stability under ambient conditions is a persistent concern, as many atomically thin vdW materials are susceptible to degradation from oxygen and moisture[13,14]. This instability is further exacerbated under optical pumping, where local heating and photo-assisted chemical reactions can accelerate material damage and degrade nonlinear performance. The resulting low damage thresholds under continuous irradiation severely constrain operating conditions and long-term device stability.

Among vdW $\chi^{(2)}$ media layered ferroelectric crystals are especially attractive because their broken inversion symmetry can support a nonlinear response across a wide thickness range, offering practical flexibility for SPDC [15–20]. In particular, the niobium oxide dihalides $NbOX_2$ (X = Cl, Br, or I) family has recently drawn attention owing to its suitability for heterostructure assembly and its exciton-enhanced second-order nonlinearity[21-26]. However, the excitonic resonance that enhances the nonlinear response also introduces non-negligible optical absorption near the pump wavelength. This absorption contributes to localized photothermal heating, ultimately leading to pump-induced damage. Therefore, encapsulation is essential to maintain both environmental stability and high nonlinearity under continuous operation.

Here, we address these challenges using ferroelectric $NbOI_2$ flakes stabilized by graphene encapsulation. Unlike the conventional h-BN encapsulation widely used for vdW materials, graphene provides not only effective environmental protection but also efficient lateral heat spreading, owing to its exceptionally high thermal conductivity[27]. This dual functionality enables enhanced heat dissipation under continuous laser irradiation and suppresses pump-assisted degradation. To realize a stable and bright vdW SPDC source, we directly compare $NbOI_2$ with graphene encapsulation, h-BN encapsulation, and no encapsulation within the same material platform. With graphene encapsulation, we obtain a record SPDC rate of $258 \pm 1.2$ Hz from $NbOI_2$ and observed a maximum $g^{(2)}(0)$ of $2816 \pm 243$. Furthermore, we generate polarization entangled photon pairs with fidelity of 94% to a maximally entangled state by stacking two crystals. Overall, our results establish graphene-encapsulated NbOI$_2$ as a stable and high-performance SPDC platform suitable for integrated quantum light generation.

**SHG-based domain characterization and graphene encapsulation**

We first examine whether a conductive graphene overlayer alters the spontaneous ferroelectric polarization of NbOI$_2$ or reduces its effective $\chi^{(2)}$ response. The SHG response is measured using a reflection-mode microscopy setup (Supplementary Fig. S1).

Fig. 1a shows the crystal structure of the NbOI$_2$ monoclinic lattice (space group C2), whose non-centrosymmetric symmetry supports a finite $\chi^{(2)}$. Fig. 1b shows an microscope image of a NbOI$_2$ flake after graphene encapsulation, confirming uniform and conformal coverage across the crystal surface. This configuration enables a direct, region by region comparison of the second harmonic generation (SHG) response from the same NbOI$_2$ flake before and after graphene encapsulation[28]. SHG mapping performed on the identical region shows that the SHG intensity changes by less than 20% after graphene encapsulation (Fig. 1c, d), indicating that the effective $\chi^{(2)}$ response is largely preserved. Despite this change in SHG strength, graphene encapsulation leads to a pronounced improvement in environmental robustness. While the unencapsulated flake degrades readily under ambient exposure and optical illumination, the graphene encapsulated NbOI$_2$ maintains a stable SHG response under the same conditions (Supplementary Fig. S2). Figs. 1e-g illustrate how graphene encapsulation not only reduces pump induced damage through heat dissipation but also protects the material from ambient conditions.

Together, these results demonstrate that graphene encapsulation preserves the ferroelectric domain integrity and $\chi^{(2)}$ functionality of NbOI$_2$ while substantially enhancing its resistance to laser and environment induced damage. This combination of nonlinear functionality and operational robustness provides a reliable foundation for high brightness SPDC under continuous excitation.

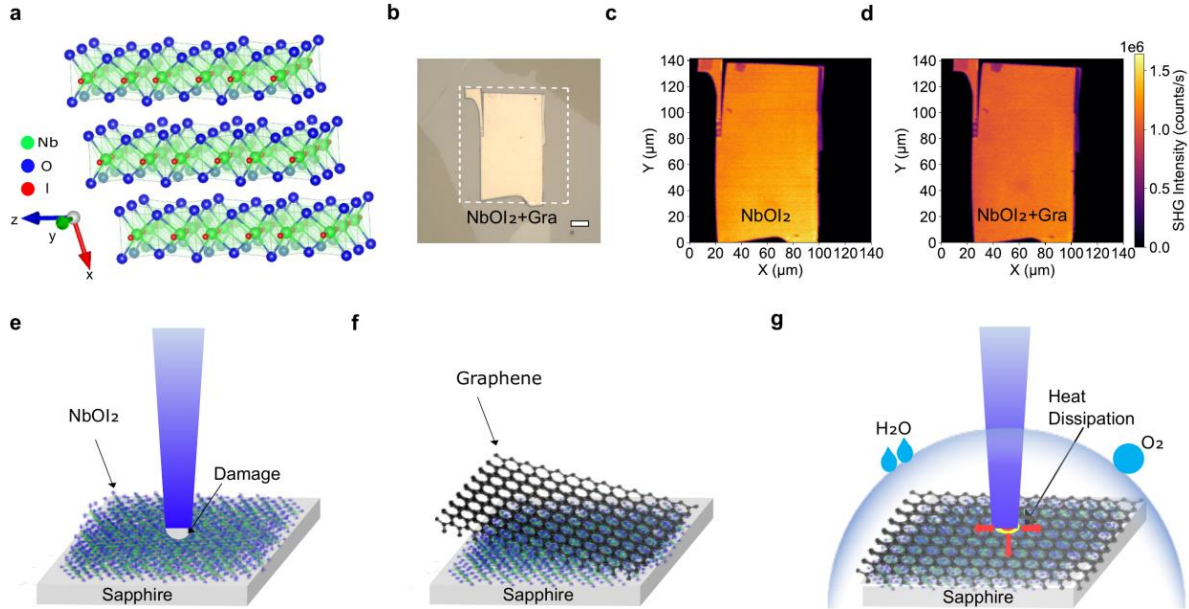

**Fig. 1| a**, Crystal structure of $NbOI_2$ showing the layered arrangement of Nb, O, and I atoms. **b**, Microscopic image of $NbOI_2$ with graphene encapsulation. The area marked with white dashed line is the imaging region for SHG mapping. The scale bar is 20 μm. **c**, SHG (Second Harmonic Generation) map showing uniform domains of $NbOI_2$. **d**, SHG map showing the SHG intensity of graphene covered $NbOI_2$ indicating the intensity is close to the no encapsulated $NbOI_2$. **e**, Schematic illustration of damage to $NbOI_2$ when the laser is applied directly without encapsulation. **f**, Schematic illustration of graphene covering $NbOI_2$. **g**, Schematic illustration of graphene protecting the flake from moisture ($H_2O$) and oxygen ($O_2$) and dissipating heat from the laser. The color bar shows the SHG Intensity (counts/sec).

## Comparison of SPDC performance in graphene, h-BN, and unencapsulated $NbOI_2$

$NbOI_2$ is known to degrade under ambient conditions, and previous work has shown that h-BN encapsulation can preserve its chemical stability[29]. However, the relatively limited thermal conductivity of h-BN [30] can restrict heat dissipation under continuous laser excitation, which may compromise the long-term operational stability of $NbOI_2$ based SPDC sources. We therefore compare graphene encapsulated, h-BN encapsulated, and unencapsulated $NbOI_2$ flakes within the same material platform to examine how different encapsulation strategies influence SPDC brightness, photon statistics, and stability under continuous laser excitation.

Fig. 2a illustrates the experimental configuration used for SPDC measurements. A continuous wave 405 nm laser is used as the pump, with its power and polarization controlled by a half-wave plate (HWP) and a polarization beam splitter (PBS). The pump beam is focused onto the sample using a 13× objective, and the generated SPDC photons are collected by a 100× objective with 0.5 NA. After spectral filtering with filter bandwidth of 10 nm to block the pump and select down converted photons centered at 810 nm, the signal is split by a beam splitter with 50% transmittivity and coupled into multimode fibres for coincidence detection. Microscopic images are acquired before and after laser exposure as shown in Fig. 2(b). Under identical irradiation conditions, the unencapsulated NbOI$_2$ flake exhibits severe damage, the h-BN encapsulated flake shows partial degradation, and the graphene encapsulated flake remains largely intact. This qualitative trend directly correlates with the SPDC performance discussed below.

For an ideal SPDC source, the coincidence rate scales linearly with pump power, the second order correlation function $g^2(0)$ decreases inversely with pump while remaining greater than 2, and the single photon count rate increases linearly with pump power[31]. Deviations from these behaviours provide a sensitive indicator of laser-induced degradation and increased scattering. The SPDC response of the unencapsulated NbOI$_2$ flake is shown in Fig. 2c-f. Schematic and white light imaging (inset) are shown in Fig. 2c (and Supplementary Fig. S3). The coincidence rate initially increases with pump power from 1mW to 40 mW but rapidly saturates and deviates from linear scaling (Fig. 2d). Upon decreasing the pump power from 40mW to 1mW, the coincidence rate does not recover, indicating irreversible damage. Correspondingly, $g^2(0)$ fails to maintain an inverse dependence on pump power during the downward power sweep (Fig. 2e). This behaviour coincides with the the single count rates in Fig. 2f, where enhanced scattering induced by optical damage leads to increased accidental coincidences and significantly suppresses the expected nonclassical statistics.

The h-BN encapsulated NbOI$_2$ flake shows partial improvement but remains unstable at higher pump powers (Fig. 2g-j). Schematic and white light imaging (inset) are shown in Fig. 2g. While the coincidence rate appears closer to linear during the initial power increase (Fig. 2h), the raw data is subjected to significant accidental coincidence counts. These accidental counts were characterized in the time-correlated single photon counting histogram data and subracted to estimate the correlated photon pair generation rate. In Fig. 2g (inset), we see a black spot, which indicates that as the power increases, h-BN starts to get damage, and due to scattering from the black spot, the single counts rapidly increases, leading to lower $g^2(0)$, as shown in Fig. 2i and j. As the pump power exceeds ~30 mW, the h-BN encapsulation is completely damaged, leaving the pump interact with NbOI$_2$ alone. In the absence of h-BN scattering, the single count rate decreases as the pump power increases. We note that after high power exposure, the SPDC response does not recover during the downward power sweep, indicating that the underlying NbOI$_2$ flake has also been damaged.

In contrast, the graphene encapsulated NbOI$_2$ flake exhibits stable SPDC behaviour across the entire pump power range (Fig. 2k-n). Schematic and white light imaging (inset) are shown in Fig. 2k. The coincidence rate remains approximately linear during both increasing and decreasing power sweeps (Fig. 2l), while $g^2(0)$ follows the expected inverse dependence on pump power without hysteresis (Fig. 2m). The single photon count rate also increases linearly and reproducibly with pump power in both directions (Fig. 2n), indicating minimal additional scattering and preserved nonlinear conversion efficiency.

Taken together, these results demonstrate that graphene encapsulation provides a uniquely effective combination of environmental protection and optical stability for vdW SPDC sources. While no encapsulation make the sample vulnerable to environment and laser damage, leading to its quick degradation, h-BN offers partial protection, but it fails under sustained high power excitation, whereas graphene enables robust, reversible, and high-brightness SPDC operation under continuous laser irradiation.

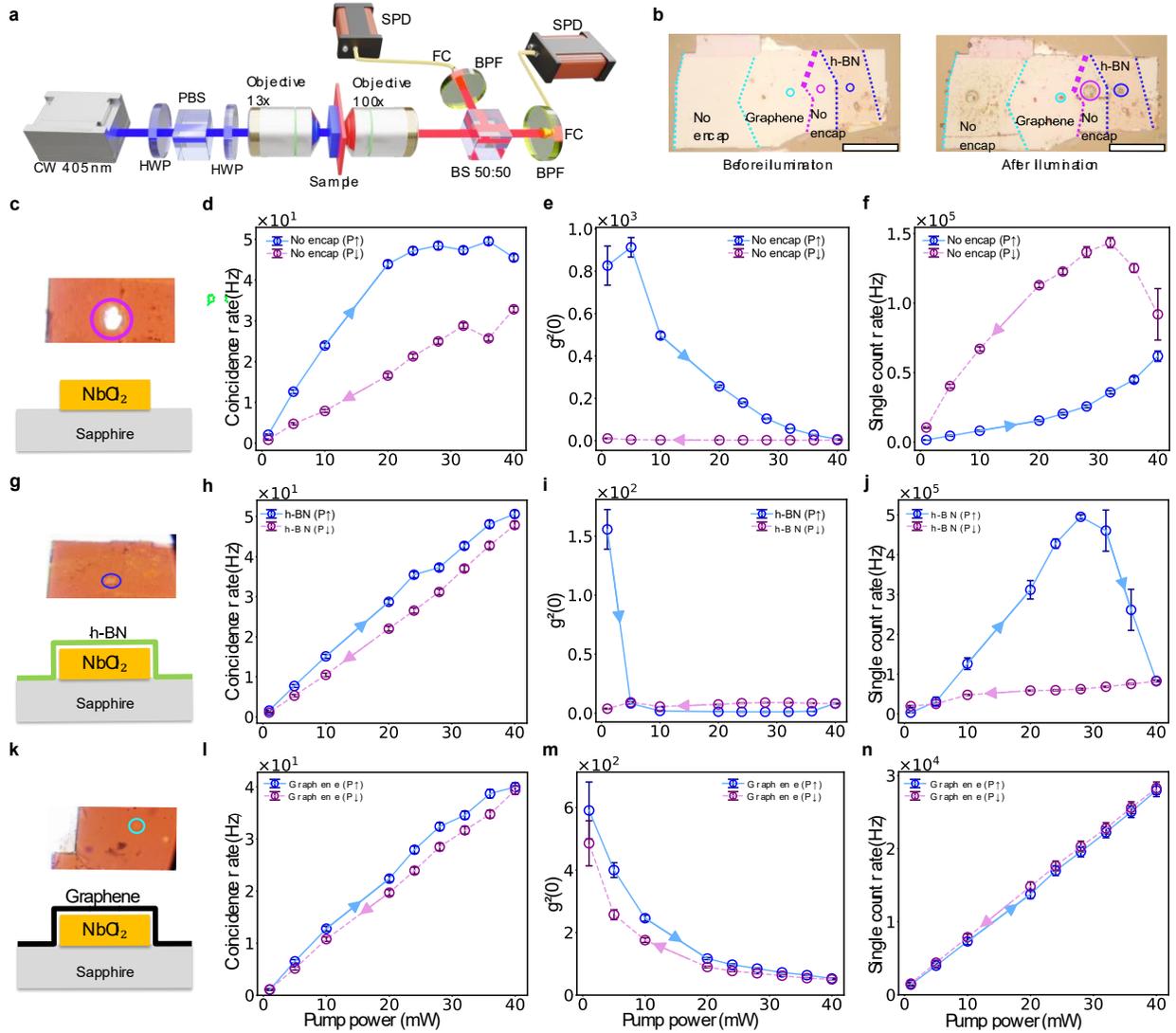

**Fig. 2| Stability of NbOI$_2$-based SPDC emission under different encapsulation conditions. a,** Experimental setup for photon-pair generation based on spontaneous parametric down-conversion (SPDC). CW, continuous-wave laser; HWP, half-wave plate; PBS, polarizing beam splitter; BS, 50:50 beam splitter; FC, fibre coupler; BPF, band-pass filter; SPD, single-photon detector. **b,** Optical microscopy images of the NbOI$_2$ flake before and after laser illumination at the indicated region. The scale bar is 50 μm. **c-f,** Bare NbOI$_2$ on sapphire: schematic illustration and optical image (inset), coincidence rate, second-order correlation at zero delay $g^{(2)}(0)$, and single-count rate as a function of pump power, showing rapid performance degradation under illumination. **g-j,** h-BN-encapsulated NbOI$_2$: corresponding schematic, optical image (inset), coincidence rate, $g^{(2)}(0)$, and single-count rate versus pump power, indicating improved stability. **k-n,** Graphene-encapsulated NbOI$_2$: corresponding schematic, optical image (inset),

coincidence rate, $g^{(2)}(0)$, and single-count rate versus pump power, demonstrating the highest robustness against laser-induced degradation.

**Enhanced SPDC performance**

After validating the stabilizing effect of graphene encapsulation on NbOI$_2$, we further optimized the SPDC performance by improving the collection optics. We used the same setup as shown in Fig. 2a, but with a collection objective of 100× magnification and 0.7 NA to improve the collection of the SPDC photons. This is because vdW material emits down converted photons in all directions, increasing NA will improve the collection efficiency, leading to higher counts. The NbOI$_2$ flake used in our experiment has a thickness of approximately 350 nm and is encapsulated with an ~8 nm graphene layer. The thickness was selected to be close to the coherence length $L_c \approx 424$nm of NbOI$_2$ to optimize the nonlinear SPDC interaction. The graphene encapsulation provides environmental protection and remains sufficiently thin to avoid significant pump absorption. Fig. 3a shows the temporal second-order correlation function of photons emitted from the NbOI$_2$ flake under 405 nm excitation. A significant peak at zero time delay with $g^{(2)}(0) \gg 2$[32] clearly indicates the generation of strongly time-correlated photon pairs. Following this, Fig. 3b shows $g^{(2)}(0)$ as a function of pump power. At low power $g^{(2)}(0)$ initially increases rapidly with increasing pump power until it reaches a maximum value of 2816 ± 243. This behaviour arises because the accidental coincidence rate is comparable to the true coincidence rate. As the pump power continues to increase, the true coincidence rate becomes dominant over accidental counts, leading to $g^{(2)}(0)$ exhibiting the expected inverse dependence on pump power[31]. Fig. 3c shows that the coincidence rate increases linearly with pump power, achieving a record high value of 258 ± 1.2 Hz for SPDC from vdW materials[33]. Assuming isotropic emission, a geometric collection efficiency of a 0.7 NA lens (~29%), a filter transmission (90%), and a detector efficiency (60%), this corresponds to an intrinsic paired photon generation rate of 10.08 ± 0.05 kHz. This estimate neglects additional optical losses and therefore represents a lower bound on the intrinsic generation rate. Finally, Fig. 3d presents the dependence of coincidence counts on pump polarization, revealing a sinusoidal dependence consistent with the anisotropic response of the NbOI$_2$ sample. The maximum coincidence count is achieved when the pump polarization along y direction is aligned with the crystallographic y-axis, making $\chi^{(2)}_{yyy}$ the dominant contribution to coincidence counts. Thus, when the pump is horizontally polarized $|H\rangle$, which is along the y-axis, we will get signal and idler photons that are both horizontally polarized (Supplementary Fig. S4). These results establish graphene encapsulated NbOI$_2$ as a robust, high brightness source of quantum photon pairs with anisotropic behaviour.

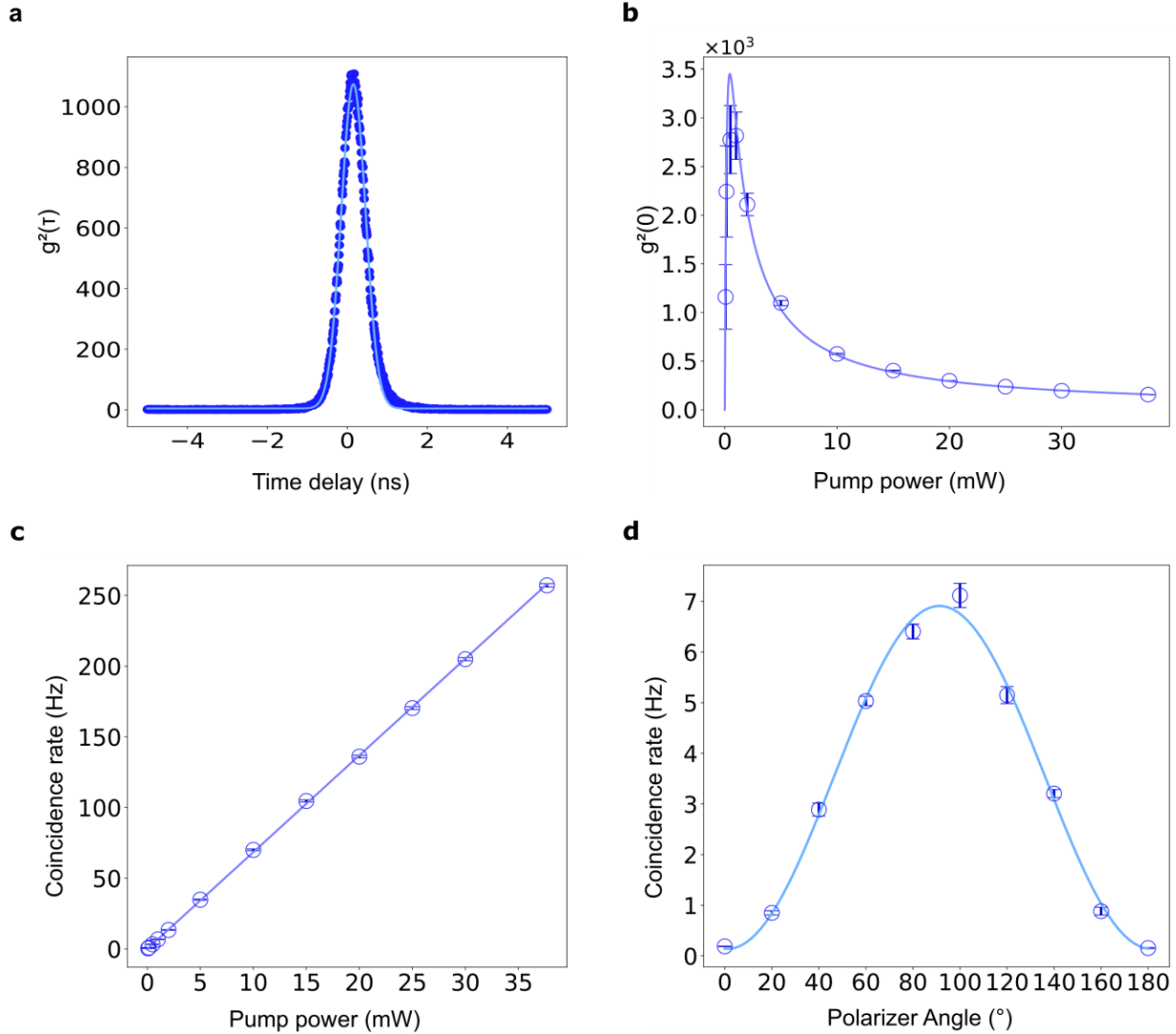

**Fig. 3| Spontaneous parametric down-conversion (SPDC) data for NbOI$_2$ of thickness 350 nm, covered with an ~8 nm graphene layer and pumped with a 405 nm continuous-wave (CW) laser. a**, Normalized second-order correlation $g^{(2)}(\tau)$ as a function of time delay $\tau$, using a bin width of 10 ps, a pump power of 10 mW, and an integration time of 40 min. **b**, $g^{(2)}(0)$ as a function of pump power, showing an inverse dependence above 0.5 mW and a nearly linear trend below this value due to accidental counts. **c**, Linear dependence of coincidence rate on pump power, with a maximum rate of 258 ± 1.2 Hz. **d**, Coincidence rate vs pump power, showing the anisotropic behaviour of the NbOI$_2$ sample.

**Polarization entanglement**

With the SPDC performance firmly established, we further demonstrate the generation of entangled photon pairs. Two photon entangled states have a broad range of applications in quantum communication[34] and quantum metrology[35]. They have been demonstrated by various approaches using bulk material like PPKTP[36], BBO[37], and also vdW material like GaP[38], MoS$_2$[39], GaSe[40], NbOCl$_2$[41, 43]. Since NbOI$_2$ has a dominant $\chi^{(2)}_{yyy}$ component, we employ a 90° stacked NbOI$_2$ heterostructure composed of a thin flake placed on top of a thicker one and encapsulated by graphene to protect against laser induced damage. The device consists of two NbOI$_2$ flakes of thickness $d_1$ and $d_2$, where the second flake is rotated by 90° relative to the first. We used thicknesses of approximately 30 nm and 207 nm for the stacked NbOI$_2$ layers (see Supplementary Section S8 for details on the thickness selection), and the heterostructure was encapsulated with a 10 nm graphene layer. This asymmetric thin-thick configuration was chosen due to the strong anisotropic polarization dependent absorption at 405 nm light in NbOI$_2$[42,43]. If layers of identical thickness were used, the first flake would absorb most of the pump, resulting in an unbalanced polarization state. Therefore, the first flake must remain thin enough to transmit sufficient pump power to the second flake, while the second flake is designed to be thick enough to generate the substantial SPDC photons from the attenuated pump. As in the characterization of SPDC photons, here we define $|H\rangle$ polarization along y axis and $|V\rangle$ polarization along $x$-axis. With the pump polarization set at θ° (polarizer angle), the pump electric field ($E_p$) is given by

$$E_p = \frac{E_0}{\sqrt{2}}\left(\sin(\theta)\, H_y + \cos(\theta)\, V_x\right)$$

where $E_0$ is the incident pump amplitude. Because the dominant tensor component in NbOI$_2$ is $\chi^{(2)}_{yyy}$ [29], the thin flake (aligned along $y$) primarily converts the $H_y$ pump component into $|HH\rangle$ photon pairs, while the $V_x$ component is largely unaffected. Conversely, at the thick flake, the $V_x$ component becomes aligned with the local $y'$ axis of the flake, where the same tensor element enables efficient generation of $|VV\rangle$ photon pairs. An input HWP was used to adjust the pump polarization to an angle of approximately 19°. This configuration yields $|HH\rangle$ and $|VV\rangle$ photon pairs with approximately equal probability.

Fig. 4 shows the quantum state tomography data for the 90° thin-thick stacked NbOI$_2$ heterostructure. For this measurement, we used a pump power of 2 mW and a coincidence window of 0.5 ns. The reconstructed real and imaginary parts of the two photon density matrix are shown in (Fig. 4a,b). We observe a fidelity of 94% with respect to the maximally entangled Bell state $(|HH\rangle + |VV\rangle)/\sqrt{2}$. These results confirm that stacked $NbOI_2$ heterostructures can serve as a reliable platform for generating high quality polarization entangled photon pairs.

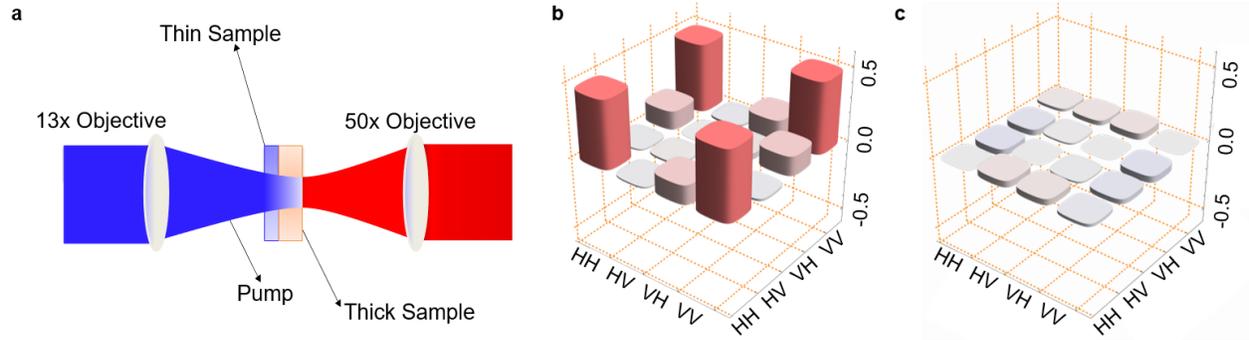

**Fig. 4| Tomography data for a 90° thin-thick stacked $NbOI_2$ heterostructure.** The generated state is $(|HH\rangle + |VV\rangle)/\sqrt{2}$. **a**, Schematic showing how the pump interacts with the thin-thick sample to generate polarization entangled photon pairs. **b,** Real part and **c,** imaginary part of the reconstructed density matrix.

## Conclusion

In conclusion, we have shown that the environmental and optical stability of vdW material-based quantum SPDC sources have a critical impact on their performance. Unencapsulated $NbOI_2$ is highly susceptible to humidity and laser-induced degradation, and hence its SPDC performance is significantly compromised. By encapsulating $NbOI_2$ with graphene, we improved not only the environmental stability but also the optical stability through efficient heat dissipation, enabling stable SPDC operation under continuous pumping. The graphene-encapsulated $NbOI_2$ devices exhibit record-high $g^{(2)}(0)$ value of $2816 \pm 243$ and coincidence rate of $258 \pm 1.2$ Hz for vdW-based SPDC sources. Furthermore, by employing a 90° stacked thin-thick $NbOI_2$ heterostructure, we realize a polarization entangled photon pair source with high fidelity.

The coincidence rate can be further improved by using thicker samples with thicknesses closer to the coherence length or by employing a collection objective with a higher numerical aperture. In addition, stacking-enabled quasi-phase matching offers a viable route toward even brighter SPDC sources[3]. Higher fidelity polarization entangled photon pairs can be achieved by controlling the stacking angle between the thin and thick layers to be as close to 90° as possible, while simultaneously optimizing the pump polarization to produce equal $|HH\rangle$ and $|VV\rangle$ counts. The nearly uniform domain for NbOI$_2$ makes the coincidence count rates mostly the same for samples with the same thickness, which will improve the scalability of NbOI$_2$-based SPDC devices. In summary, these results position NbOI$_2$ as a promising 2D platform for scalable on-chip quantum photonic sources.

## Methods

### Device preparation and transfer

NbOI$_2$ crystals were purchased from HQ Graphene. The flake thicknesses were measured using a KLA-Tencor P-16 profiler with a scan range up to 200 mm and a vertical range up to 110 μm. The crystals were exfoliated using the Scotch tape method onto silicon wafers with a SiO$_2$ layer. A transfer stage and capillary force assisted transfer protocol were used to place selected flakes onto sapphire substrates purchased from 2D Semiconductors [44,45].

For the encapsulation comparison, we used NbOI$_2$ flakes with thicknesses of approximately 280 nm, graphene thicknesses of 7 nm, and h-BN thicknesses of 15 nm. The choice of flake and encapsulation thickness was not strictly controlled; instead, flakes were selected such that the measured coincidence counts were as similar as possible across different encapsulation conditions. This facilitated a direct comparison of the encapsulation effects. For the coincidence rate measurements reported in Fig. 3, we used an NbOI$_2$ flake of thickness 350 nm with an ~8 nm graphene encapsulation. For the polarization entanglement experiments, we used a thick NbOI$_2$ flake of thickness 207 nm and a thin flake of 30 nm, both covered by a graphene layer of ~10 nm thickness and pumped with an input power of 2 mW.

### Parametric generation of quantum photon pairs measurements

For SPDC generation, we used an OBIS LX SF 405 nm laser. A UV objective lens (13×, 0.13 NA) was used to focus the pump onto the sample. For the encapsulation comparison measurements, SPDC photons were collected with a 100×, 0.5 NA Mitutoyo Plan Apo NIR infinity-corrected objective (480-1800 nm). For the high-count-rate measurements, we used a 100× long-working-distance Mitutoyo Plan Apo objective with 0.7 NA optimized for 400-655 nm. For the polarization-entanglement experiments, we reverted to the 100×, 0.5 NA Mitutoyo Plan Apo NIR infinity-corrected objective (480-1800 nm) for collection. A BS005 50:50 non-polarizing beam splitter (BS) cube (700-1100 nm) and an FBH810-10 hard-coated band-pass filter (Thorlabs) were used to filter and route the photons to the detectors. A 10×, 0.25 NA microscope objective (Newport) was used as the output collection objective in our measurements.

**Second harmonic generation measurements**

Second-harmonic generation (SHG) is a second-order nonlinear optical process in which an input field $E(\omega)$ induces a nonlinear polarization $P(2\omega) = \varepsilon_0 \chi^{(2)} E^2(\omega)$, producing radiation at $2\omega$. In our measurements, a pump at $\omega = 1030$nm was focused onto NbOI$_2$, generating SHG at 515 nm. Regions lacking $\chi^{(2)}$ do not produce a measurable $P(2\omega)$ and therefore show negligible SHG signal, whereas regions with nonzero $\chi^{(2)}$ exhibit strong SHG intensity, indicating efficient frequency doubling. The optical setup follows our previous report[46], and further details are provided in Supplementary Fig. 1.

**Data availability**

All data that support the findings of this study are available in the main text, figures, and Supplementary Information. They are also available from the corresponding author upon reasonable request.

**Code availability**

All code that is used in the main text and Supplementary Information are available from the corresponding author upon reasonable request.

**Acknowledgments**

This work was supported by the Agency for Science, Technology and Research (A*STAR) under MTC YIRG Grant No. M23M7c0129, Career Development Fund-Seed Projects Grant No. 222D800038, A*STAR Quantum Innovation Centre (Q. InC) SRTT. This work is also supported by the national Research Foundation, Singapore through the National Quantum Office, hosted in A*STAR, under its Centre for Quantum Technologies Funding Initiative (S24Q2d0009).

**Author contributions**

Y.C, X.M., S.M.A and J.Z conceived the idea. Y. C., X.M., S.M.A, P.K.L. supervised this project. M.J. and Y.X fabricated the devices; M.J. and Y.C. built the SPDC setup; M.-T. J built the SHG setup and characterized the SHG imaging. M.J., X.M. and Y.C. wrote the draft and all authors revised and approved the manuscript.

**Competing interests**

Authors declare that they have no competing interests.

**Reference**

1. Anwar, A., Perumangatt, C., Steinlechner, F., Jennewein, T. & Ling, A. Entangled photon-pair sources based on three-wave mixing in bulk crystals. *Rev. Sci. Instrum.* **92**, 041101 (2021).
2. Lin, K., Yao, G., Shao, J. *et al.* Nonlinear phase-matched van der Waals crystals integrated on optical fibres. *Nat. Mater.* (2026).
3. Trovatello, C. et al. Quasi-phase-matched up- and down-conversion in periodically poled layered semiconductors. *Nat. Photon.* **19**, 291-299 (2025).
4. Konacki, M., Torres, G., Jha, S. & Sasselov, D. D. Long-distance teleportation of qubits at telecommunication wavelengths. *Nature* **421**, 507-509 (2003).
5. Knill, E., Laflamme, R. & Milburn, G. J. A scheme for efficient quantum computation with linear optics. *Nature* **409**, 46-52 (2001).
6. Tan, S.-H. et al. Quantum illumination with Gaussian states. *Phys. Rev. Lett.* **101**, 253601 (2008).
7. Weihs, G., Jennewein, T., Simon, C., Weinfurter, H. & Zeilinger, A. Violation of Bell's inequality under strict Einstein locality conditions. *Phys. Rev. Lett.* **81**, 5039-5043 (1998).
8. Xiao, J. et al. Intrinsic two-dimensional ferroelectricity with dipole locking. *Phys. Rev. Lett.* **120**, 227601 (2018).
9. Somvanshi, D. & Jit, S. Advances in 2D materials based mixed-dimensional heterostructures photodetectors: Present status and challenges. *Materials Science in Semiconductor Processing.* **164** (2023).
10. Geim, A. K. & Grigorieva, I. V. Van der Waals heterostructures. *Nature.* **499**, 419-425 (2013).
11. Li, M. Y., Chen, C. H., Shi, Y. & Li, L. J. Heterostructures based on two-dimensional layered materials and their potential applications. *Materials Today.* **19**, 322-335 (2016).
12. Lim, H., Yoon, S. I., Kim, G., Jang, A. R. & Shin, H. S. Stacking of two-dimensional materials in lateral and vertical directions. *Chemistry of Materials* **26**, 4891-4903 (2014).
13. Zhao, Q. et al. Toward air stability of thin GaSe devices: avoiding environmental and laser-induced degradation by encapsulation. *Advanced Functional Materials* **28**, 1805304 (2018).
14. Kang, J. *et al.* Layered $NbOCl_2$ kinetic degradation mechanism and improved second-order nonlinear optical responses. *Mater. Adv.* **6**, 954-962 (2025).
15. Li, C. *et al.* Second harmonic generation from a single plasmonic nanorod strongly coupled to a $WSe_2$ monolayer. *Nano Lett.* **21**, 1599-1605 (2021).
16. Yu, H., Talukdar, D., Xu, W., Khurgin, J. B. & Xiong, Q. Charge-Induced Second-Harmonic Generation in Bilayer $WSe_2$. *Nano Lett.* **15**, 5653-5657 (2015).

17. Puri, S. et al. Substrate interference and strain in the second-harmonic generation from MoSe$_2$ monolayers. *Nano Lett.* **24**, 13061-13067 (2024).
18. Chen, H. *et al.* Enhanced second-harmonic generation from two-dimensional MoSe$_2$ on a silicon waveguide. *Light Sci. Appl.* **6**, e17060 (2017).
19. Janisch, C. et al. Extraordinary second harmonic generation in tungsten disulfide monolayers. *Sci. Rep.* **4**, 5530 (2014).
20. Malard, L. M., Alencar, T. V., Barboza, A. P. M., Mak, K. F. & de Paula, A. M. Observation of intense second harmonic generation from MoS$_2$ atomic crystals. *Phys. Rev. B* **87**, 201401 (2013).
21. Guo, Q. *et al.* Ultrathin quantum light source with van der Waals NbOCl$_2$ crystal. *Nature* **613**, 53-59 (2023).
22. Abdelwahab, I. *et al.* Giant second-harmonic generation in ferroelectric NbOI$_2$. *Nat. Photonics* **16**, 644-650 (2022).
23. Fu, J. et al. Emission dipole and pressure-driven tunability of second harmonic generation in vdWs ferroelectric NbOI$_2$. *Adv. Funct. Mater.* **34**, 2308207(2023)
24. Chen, W. et al. Extraordinary enhancement of nonlinear optical interaction in NbOBr$_2$ microcavities. *Adv. Mater.* **36**, 2400858 (2024).
25. Bai, L. *et al.* Photonic Crystal Defect Cavities Enable Air-Stable and Enhanced SHG from NbOCl$_2$. *Adv. Opt. Mater.* **13**, 30 (2025).
26. Xuan, F., Lai, M., Wu, Y. & Quek, S. Y. Exciton-enhanced spontaneous parametric down-conversion in two-dimensional crystals. *Phys. Rev. Lett.* **132**, 246902 (2024).
27. Balandin, A. A. *et al.* Superior thermal conductivity of single-layer graphene. *Nano Lett.* **8**, 902-907 (2008).
28. Ye, L. et al. Manipulation of nonlinear optical responses in layered ferroelectric niobium oxide dihalides. *Nat. Commun.* **14**, 5911 (2023).
29. Yan, Q. *et al.* Ambient Degradation Anisotropy and Mechanism of van der Waals Ferroelectric NbOI$_2$. *ACS Appl. Mater. Interfaces* **16**, 9051-9059 (2024).
30. Cai, Q. et al. High thermal conductivity of high-quality monolayer boron nitride and its thermal expansion. *Sci. Adv.* **5**, eaav0129 (2019).
31. Wang, Y., Jöns, K. D. & Sun, Z. Integrated photon-pair sources with nonlinear optics. *Applied Physics Reviews*. **8** (2021).
32. Ding, D.-S., Zhou, Z.-Y., Shi, B.-S., Zou, X.-B. & Guo, G.-C. Generation of non-classical correlated photon pairs via a ladder-type atomic configuration: theory and experiment. *Opt. Express* **20**, 11433-11444 (2012).
33. Liang, H. et al. Tunable polarization entangled photon-pair source in rhombohedral boron nitride. *Sci. Adv.* **11**, eadt3710 (2025).
34. Piveteau, A. et al. Entanglement-assisted quantum communication with simple measurements. *Nat. Commun.* **13**, 7878 (2022).

35. Krischek, R. et al. Useful multiparticle entanglement and sub-shot-noise sensitivity in experimental phase estimation. *Phys. Rev. Lett.* **107**, 080504 (2011).
36. Lee, S. M., Kim, H., Cha, M. & Moon, H. S. Polarization-entangled photon-pair source obtained via type-II non-collinear SPDC process with PPKTP crystal. *Opt. Express* **24**, 2941 (2016).
37. Villar, A., Lohrmann, A. & Ling, A. Experimental entangled photon pair generation using crystals with parallel optical axes. *Opt. Express* **26**, 12396 (2018).
38. Sultanov, V., Santiago-Cruz, T. & Chekhova, M. V. Flat-optics generation of broadband photon pairs with tunable polarization entanglement. *Opt. Lett.* **47**, 3872 (2022).
39. Weissflog, M. A. et al. A tunable transition metal dichalcogenide entangled photon-pair source. *Nat. Commun.* **15**, 7600 (2024).
40. Lu, Z. *et al.* Counter-propagating entangled photon pairs from monolayer GaSe. *Nat. Commun.* **16**, 9616 (2025).
41. Guo, Q. et al. Polarization entanglement enabled by orthogonally stacked van der Waals $NbOCl_2$ crystals. *Nat. Commun.* **15**, 10461 (2024).
42. Fang, Y., Wang, F., Wang, R., Zhai, T. & Huang, F. 2D $NbOI_2$: a chiral semiconductor with highly in-plane anisotropic electrical and optical properties. *Adv. Mater.* **33**, 2101505 (2021).
43. Mortazavi, B., Shahrokhi, M., Javvaji, B., Shapeev, A. V. & Zhuang, X. Highly anisotropic mechanical and optical properties of 2D $NbOX_2$ (X = Cl, Br, I) revealed by first principle. *Nanotechnology* **33**, 27 (2022).
44. Ma, X. *et al.* Capillary-Force-Assisted Clean-Stamp Transfer of Two-Dimensional Materials. *Nano Lett.* **17**, 6961-6967 (2017).
45. Huang, G. *et al.* Transfer and beyond: emerging strategies and trends in two-dimensional material device fabrication. *Chemical Society Reviews.* (2026).
46. Aletheia, W., Jiang, M., Cao, M., Huang, S., Lourembam, J., Ma, X., Duan, R. & Liu, Z. Nonlinear optical processes in 2D Cairo pentagonal palladium phosphide sulfide. *Nano Research.* (2026).